\documentclass[conference]{IEEEtran}

\usepackage[T1]{fontenc}
\usepackage{subcaption}
\usepackage{color}
\usepackage{varioref}
\usepackage{textcomp}
\usepackage{amsthm}
\usepackage{amsmath}
\usepackage{amssymb}
\usepackage{graphicx}
\usepackage{setspace}
\usepackage{esint}
\usepackage{algorithm}
\usepackage{algpseudocode}
\usepackage{pifont}
\usepackage{wrapfig}
\usepackage{multirow}
\usepackage{tabu}
\usepackage{amsmath}

\usepackage{enumitem}
\usepackage{cite}
\usepackage{cleveref}
\makeatletter

\newcommand{\Rmnum}[1]{\expandafter\@slowromancap\romannumeral #1@}
\makeatother

\newtheorem{remark}{Remark}

\theoremstyle{definition}

\providecommand{\propositionname}{Proposition}

\ifCLASSINFOpdf

\else

\fi

\usepackage[T1]{fontenc}
\usepackage{etoolbox}

\makeatletter
\patchcmd{\maketitle}{\@fnsymbol}{\@alph}{}{}  
\makeatother

\title{Collaborative Machine Learning at the Wireless Edge with Blind Transmitters}
\author{\IEEEauthorblockN{Mohammad Mohammadi Amiri\IEEEauthorrefmark{1}, Tolga M. Duman\IEEEauthorrefmark{2}, Deniz G\"und\"uz\IEEEauthorrefmark{1}}
\IEEEauthorblockA{\IEEEauthorrefmark{1}Electrical and Electronic Engineering Department, Imperial College London, London SW7 2BT, U.K.}
\IEEEauthorblockA{\IEEEauthorrefmark{2}Department of Electrical and Electronics Engineering, Bilkent University, Ankara 06800, Turkey}
Email: \IEEEauthorrefmark{1}\{m.mohammadi-amiri15, d.gunduz\}@imperial.ac.uk, \IEEEauthorrefmark{2}duman@ee.bilkent.edu.tr}

\date{}

\begin{document}
 
\maketitle

\begin{abstract}
We study wireless collaborative machine learning (ML), where mobile edge devices, each with its own dataset, carry out distributed stochastic gradient descent (DSGD) over-the-air with the help of a wireless access point acting as the parameter server (PS). At each iteration of the DSGD algorithm wireless devices compute gradient estimates with their local datasets, and send them to the PS over a wireless fading multiple access channel (MAC). Motivated by the additive nature of the wireless MAC, we propose an analog DSGD scheme, in which the devices transmit scaled versions of their gradient estimates in an uncoded fashion. We assume that the channel state information (CSI) is available only at the PS. We instead allow the PS to employ multiple antennas to alleviate the destructive fading effect, which cannot be cancelled by the transmitters due to the lack of CSI. Theoretical analysis indicates that, with the proposed DSGD scheme, increasing the number of PS antennas mitigates the fading effect, and, in the limit, the effects of fading and noise disappear, and the PS receives aligned signals used to update the model parameter. The theoretical results are then corroborated with the experimental ones.  
\end{abstract}


\section{Introduction}\label{SecIntro}

With the growing prevalence of Internet of things (IoT) devices, constantly collecting information about various physical phenomena, and the growth in the number and processing capability of mobile edge devices (phones, tablets, smart watches and activity monitors), there is a growing interest in enabling distributed machine learning (ML) to learn from data distributed across mobile devices. Centralized ML techniques are often developed, assuming that the datasets are offloaded to a central processor. In the case of wireless edge devices, centralized ML techniques are not desirable, since offloading such massive amounts of data to a central cloud may be too costly in terms of both energy and privacy. 

In many ML problems, the goal is to minimize a loss function, $F \left( \boldsymbol{\theta} \right)$, where $\boldsymbol{\theta} \in \mathbb{R}^d$ captures the model parameters to be optimized. The loss function $F \left( \boldsymbol{\theta} \right)$ represents the average of empirical loss functions computed at different data samples with respect to model parameter $\boldsymbol{\theta}$, $F \left( \boldsymbol{\theta} \right) = \frac{1}{\left| \mathcal{B} \right|} \sum\nolimits_{\boldsymbol{u} \in \mathcal{B}} f \left(\boldsymbol{\theta}, \boldsymbol{u} \right)$, where $\mathcal{B}$ is the set of available data points, and $\boldsymbol{u}$ represents a data sample and its label. 

We assume that an iterative stochastic gradient descent (SGD) algorithm is used to minimize the loss function $F \left( \boldsymbol{\theta} \right)$, in which the model parameter vector at iteration $t$, $\boldsymbol{\theta}_t$, is updated according to the stochastic gradient $\boldsymbol{g} \left( \boldsymbol{\theta}_t \right)$. SGD allows parallelization across multiple mobile devices. In distributed SGD (DSGD), devices process data locally with respect to a globally consistent parameter vector, and send their gradient estimates to the parameter server (PS). To be more precise, at iteration $t$, device $m$ computes the gradient estimate $\boldsymbol{g}_m \left( \boldsymbol{\theta}_t \right) \triangleq \frac{1}{\left| \mathcal{B}_{m} \right|} \sum\nolimits_{\boldsymbol{u} \in \mathcal{B}_{m}} \nabla f \left(\boldsymbol{\theta}_t, \boldsymbol{u} \right)$ with respect to its local dataset $\mathcal{B}_m$ and model parameter $\boldsymbol{\theta}_t$, and sends the result to the PS. Having $M$ devices in the system, the PS updates the model parameter vector according to     
\begin{align}\label{ParallelSGDModelUpdate}
\boldsymbol{\theta}_{t+1} =\boldsymbol{\theta}_{t} - \eta_t \frac{1}{M} \sum\nolimits_{m=1}^{M} \boldsymbol{g}_m \left(\boldsymbol{\theta}_{t} \right),   
\end{align}
where $\eta_t$ denotes the learning rate at iteration $t$, and shares the result with the devices for the computations at the following iterations. Although parallelism reduces the computation load at each device, communication from the devices to the PS becomes the main performance bottleneck \cite{DCAlistarhQSGD,DCOneBitQuan,DCLimitedPrecisionGupta,ScalableDNNStorm,DCMohammadDenizScheduling}, particularly for wireless edge learning due to limited bandwidth and power.

Several architectures have been proposed in recent years to employ computational capabilities of edge devices, and train an ML model collaboratively with the help of a remote PS. However, these works ignore the physical characteristics of the communication channel from the devices to the PS, and consider interference-and-error-free links with a fixed capacity, which is hard to guarantee in most wireless environments.



Collaborative ML taking into account the physical layer channel characteristics has recently been studied in \cite{MohammadDenizDSGDCS, KaibinParallelWork, YangFedLearOverAirComp, MohammadDenizSpawc19}. These works consider a wireless multiple access channel (MAC) from the edge devices to the PS, and propose over-the-air computation to average gradient vectors or estimated model parameters at the PS. In \cite{MohammadDenizDSGDCS}  the authors focus on bandwidth efficient learning, and employ gradient sparsification followed by linear projection to design a communication efficient DSGD algorithm. This scheme has been extended to the fading MAC model in \cite{MohammadDenizSpawc19}. Distributed ML  over a wireless fading MAC is studied in \cite{KaibinParallelWork}, where the wireless devices employ power allocation with perfect channel state information (CSI) to align the received signals at the PS. A single-input multiple-output (SIMO) wireless fading MAC is studied in \cite{YangFedLearOverAirComp}, where a beamforming technique is designed to maximize the number of devices participating in each iteration, while keeping the quality of the received signal at the PS above the specified threshold level.

Our goal in this paper is to enable distributed learning over a wireless fading MAC, while removing the requirement of CSI at the transmitters (CSIT). This will be achieved by employing multiple antennas at the PS. Similarly to \cite{MohammadDenizDSGDCS, KaibinParallelWork, YangFedLearOverAirComp, MohammadDenizSpawc19} we considering uncoded transmission of gradient estimates and over-the-air computation.
We design a receive beamformer at the PS in order to mitigate the fading effect and align the desired signals. We analytically show that the proposed scheme alleviates the destructive effects of interference and noise terms at the PS thanks to the utilization of multiple antennas, and, in the limit, due to channel hardening, it boils down to a deterministic channel with identical gains from all the devices. This result is validated by numerical experiments, where we investigate the impact of the number of antennas on the performance of the proposed scheme with no CSIT. It is worth noting that the CSI requirements of over-the-air computation with a multi-antenna receiver was also studied in \cite{ComOverAirNoCSIT}. The authors proposed a scheme that encodes the information on the energy of the transmitter signals, and hence, limited only to positive values, but requires CSI neither at the transmitters nor at the PS. Performance of this no-CSI scheme for DSGD will be studied in the extended version of this paper. 


\textit{Notations}: $\mathbb{R}$ and $\mathbb{C}$ represent the sets of real and complex values, respectively. We denote entry-wise complex conjugate of vector $\boldsymbol{x}$ by $\left( \boldsymbol{x} \right)^*$, and ${\rm{Re}} \{ \boldsymbol{x} \}$ and ${\rm{Im}} \{ \boldsymbol{x} \}$ return entry-wise real and imaginary components of $\boldsymbol{x}$, respectively. For $\boldsymbol{x}$ and $\boldsymbol{y}$ with the same dimension, $\boldsymbol{x} \cdot \boldsymbol{y}$ returns their inner product. We denote a zero-mean normal distribution with variance $\sigma^2$ by $\mathcal{N} \left( 0,\sigma^2 \right)$, and $\mathcal{C N} \left( 0,\sigma^2 \right)$ represents a circularly symmetric complex normal distribution with real and imaginary terms each distributed according to $\mathcal{N} \left( 0,\sigma^2 / 2 \right)$. We let $[i] \triangleq \{ 1, \dots, i \}$. We denote the cardinality of set $\cal X$ by $\left| \mathcal{X} \right|$, and $l_2$ norm of vector $\boldsymbol{x}$ by $\left\| \boldsymbol{x} \right\|_2$.

\section{System Model}\label{SecProbFormul}

We consider $M$ devices, where device $m$ has access to a local dataset $\mathcal{B}_m$, and employs SGD to compute the gradient estimate $\boldsymbol{g}_m \left( \boldsymbol{\theta}_t \right) \in \mathbb{R}^d$ at iteration $t$, $m \in [M]$. These local gradient estimates are transmitted to the PS, equipped with $K$ antennas, through a wireless shared medium. The PS updates the model parameter based on its received signal, and shares it with all the devices over an error-free shared link, so that all the devices have a globally consistent model parameter.

We model the shared wireless channel from the edge devices to the PS as a wireless fading MAC, where OFDM is used to divide the available bandwidth into $s$ subchannels, $s \le d$ (in practice, we typically have $s \ll d$). We assume that $N$ OFDM symbols can be transmitted over each subchannel at each iteration of DSGD algorithm. The received vector corresponding to the $n$-th OFDM symbol in iteration $t$ at the $k$-th antenna of the PS is given by
\begin{align}\label{ReceivedVectorPSGenAntennak}
\boldsymbol{y}^n_k (t) = \sum\nolimits_{m = 1}^{M} \boldsymbol{h}^n_{m,k} (t) \cdot \boldsymbol{x}^n_{m} (t) + \boldsymbol{z}^n_k (t), \quad \mbox{$k \in [K]$},
\end{align}
where $\boldsymbol{x}^n_{m} (t)$ is the $n$-th symbol of dimension $s$ transmitted by the $m$-th device, $\boldsymbol{h}^n_{m,k} (t) \in \mathbb{C}^s$ denotes the vector of channel gains from device $m$ to the $k$-th PS antenna, $m \in [M]$, and $\boldsymbol{z}^n_{k} (t) \in \mathbb{C}^s$ represents the circularly symmetric complex white Gaussian noise at the $k$-th antenna of the PS, $n \in [N]$. The $i$-th entry of channel vector $\boldsymbol{h}^n_{m,k} (t)$, denoted by $h^n_{m,k,i} (t)$, is distributed according to $\mathcal{C N} \left( 0, \sigma_h^2 \right)$, $i \in [s]$, and different entries of $\boldsymbol{h}^n_{m,k} (t)$ can be correlated, while the channel gains are assumed to be independent and identically distributed (i.i.d.) across PS antennas, OFDM symbols, and wireless devices, $k \in [K]$, $n \in [N]$, $m \in [M]$. Similarly, different entries of noise vector $\boldsymbol{z}^n_k (t)$ can be correlated, and its $i$-th entry, denoted by $z^n_{k,i} (t)$, distributed according to $\mathcal{C N} \left( 0, \sigma_z^2 \right)$, $i \in [s]$, $k \in [K]$, $n \in [N]$. Noise vectors are also assumed to be i.i.d. across PS antennas and OFDM symbols. We consider the following average power constraint imposed at each wireless device assuming a total of $T$ iterations of the DSGD algorithm:       
\begin{align}\label{AvePowerConsGen}
\frac{1}{NT} \sum\nolimits_{t=1}^{T} \sum\nolimits_{n=1}^{N} \mathbb{E} \left[ ||\boldsymbol{x}^n_{m} (t)||^2_2 \right] \le \bar{P}, \quad \forall m \in [M],
\end{align}
where the expectation is taken with respect to the randomness of the communication channel.

We assume that the PS has perfect CSI, while there is no CSI at the wireless devices. At each iteration, the goal at the PS is to estimate the average of the gradient estimates, $\frac{1}{M} \sum\nolimits_{m=1}^{M} \boldsymbol{g}_m \left(\boldsymbol{\theta}_{t} \right)$, denoted by $\hat{\boldsymbol{g}} \left(\boldsymbol{\theta}_{t} \right)$, and update the model parameter as in \eqref{ParallelSGDModelUpdate} at the end of each iteration based on the received symbols $\boldsymbol{y}^1_k (t), \ldots, \boldsymbol{y}^N_k (t)$, $\forall k$, and its knowledge of the CSI $\boldsymbol{h}^n_{m,k} (t)$, $\forall k, n, m$.

We note that the PS is interested in the average of the gradient estimates computed by the devices rather than each individual estimate. Motivated by the additive nature of the wireless MAC, we consider an analog approach similarly to \cite{MohammadDenizDSGDCS,KaibinParallelWork,YangFedLearOverAirComp,MohammadDenizSpawc19}, where the devices transmit their gradient estimates simultaneously without employing any channel coding.

\section{Analog DSGD without CSIT}\label{SecPeoposedAnalog}


At iteration $t$ of DSGD, device $m$ transmits its gradient estimate ${\boldsymbol{g}}_m \left(\boldsymbol{\theta}_{t} \right) \in \mathbb{R}^{d}$ over $N = \left\lceil {d/2s} \right\rceil$ OFDM symbols across $s$ subchannels in an uncoded manner, $m \in [M]$. We denote the $i$-th entry of ${\boldsymbol{g}}_m \left(\boldsymbol{\theta}_{t} \right)$ by $g_{m,i} \left(\boldsymbol{\theta}_{t} \right)$, $i \in [d]$, and define, for $n \in [N]$, $m \in [M]$,
\begin{subequations}
\label{gnmESADSGDDef}
\begin{align}\label{gnmRealESADSGDDef}
{\boldsymbol{g}}^n_{m, {\rm{re}}} \left(\boldsymbol{\theta}_{t} \right)& \triangleq [ g_{m,2(n-1)s+1} \left(\boldsymbol{\theta}_{t} \right), \cdots, g_{m,(2n-1)s} \left(\boldsymbol{\theta}_{t} \right)]^T,\\
{\boldsymbol{g}}^n_{m, {\rm{im}}} \left(\boldsymbol{\theta}_{t} \right)& \triangleq [ g_{m,(2n-1)s+1} \left(\boldsymbol{\theta}_{t} \right), \cdots, g_{m,2ns} \left(\boldsymbol{\theta}_{t} \right)]^T,
\label{gnmImagESADSGDDef}\\
{\boldsymbol{g}}^n_{m} \left(\boldsymbol{\theta}_{t} \right) & \triangleq {\boldsymbol{g}}^n_{m, {\rm{re}}} \left(\boldsymbol{\theta}_{t} \right) + j {\boldsymbol{g}}^n_{m, {\rm{im}}} \left(\boldsymbol{\theta}_{t} \right),
\label{gnmRealImagESADSGDDef}
\end{align}
\end{subequations}
where $j \triangleq \sqrt{-1}$, and we zero-pad ${\boldsymbol{g}}_m \left(\boldsymbol{\theta}_{t} \right)$ to have length $2sN$. The $i$-th entry of ${\boldsymbol{g}}^n_{m} \left(\boldsymbol{\theta}_{t} \right)$ is then given by
\begin{align}\label{ithgmn}
g^n_{m,i} \left(\boldsymbol{\theta}_{t} \right) = g_{m,2(n-1)s+i} & \left(\boldsymbol{\theta}_{t}  \right) + j g_{m,(2n-1)s+i} \left(\boldsymbol{\theta}_{t} \right), \nonumber\\
& \mbox{for $i \in [s]$, $n \in [N]$, $m \in [M]$}.
\end{align}
According to \eqref{gnmESADSGDDef}, we have
\begin{align}\label{gmwrtgmnReIm}
{\boldsymbol{g}}_m \left(\boldsymbol{\theta}_{t} \right) = & \big[ {\boldsymbol{g}}^1_{m, {\rm{re}}} \left(\boldsymbol{\theta}_{t} \right), {\boldsymbol{g}}^1_{m, {\rm{im}}} \left(\boldsymbol{\theta}_{t} \right), \cdots, \big. \nonumber\\
& \qquad \qquad \qquad \quad \big. {\boldsymbol{g}}^N_{m, {\rm{re}}} \left(\boldsymbol{\theta}_{t} \right), {\boldsymbol{g}}^N_{m, {\rm{im}}} \left(\boldsymbol{\theta}_{t} \right) \big]^T,
\end{align}
with $N = \left\lceil {d/2s} \right\rceil$. At the $n$-th OFDM symbol of iteration $t$, device $m$ sends 
\begin{align}\label{workermSends}
\boldsymbol{x}^n_{m} (t) = \alpha_t \boldsymbol{g}^n_{m} (t), \quad n \in [N], m \in [M].
\end{align}
Accordingly, the average transmit power depends on $\alpha_t$, and is evaluated as follows: 
\begin{align}\label{AvePowerConsWorkerm}
\frac{1}{NT} \sum\nolimits_{t=1}^{T} \alpha_t^2 \sum\nolimits_{n=1}^{N} ||\boldsymbol{g}^n_{m} (t)||^2_2 \le \bar{P}. 
\end{align}

The PS observes the following signal at its $k$-th antenna, for $k \in [K], n \in [N]$:
\begin{align}\label{ReceivedVectorPSScheAntennak}
\boldsymbol{y}^n_k (t) = \alpha_t \sum\nolimits_{m = 1}^{M} \boldsymbol{h}^n_{m,k} (t) \cdot \boldsymbol{g}^n_{m} (t) + \boldsymbol{z}^n_k (t). 
\end{align}
Having known the CSI, the PS combines the signals at different antennas in the following form:
\begin{align}\label{ReceivedVectorPSScheCombAntennas}
\boldsymbol{y}^n (t) \triangleq \frac{1}{K} \sum\nolimits_{k=1}^{K} \left( \sum\nolimits_{m = 1}^{M} \boldsymbol{h}^n_{m,k} (t) \right)^{*} \cdot \boldsymbol{y}^n_k (t), 
\end{align}
whose $i$-th entry is given by
\begin{align}\label{ReceivedVectorPSScheCombAntennasith}
y^n_i (t) = \frac{1}{K} \sum\nolimits_{k=1}^{K}  \sum\nolimits_{m = 1}^{M} \left( {h}^n_{m,k,i} (t) \right)^{*} {y}^n_{k,i} (t), 
\end{align}
where ${y}^n_{k,i} (t)$ denotes the $i$-th entry of $\boldsymbol{y}^n_{k,i} (t)$, $i \in [s]$, $n \in [N]$. By substituting ${y}^n_{k,i} (t)$, given in \eqref{ReceivedVectorPSScheAntennak}, it follows that
\begin{align}\label{ReceivedVectorPSScheCombAntennasReWrith}
&{y}^n_i (t) = \underbrace{\alpha_t \sum\limits_{m=1}^{M} \left( \frac{1}{K} \sum\limits_{k=1}^{K} \left| {h}^n_{m,k,i} (t) \right|^2 \right)  {g}^n_{m,i} (\boldsymbol{\theta}_t)}_{\text{\normalfont signal term}} \nonumber\\
& \quad + \underbrace{\frac{\alpha_t}{K} \sum\limits_{k=1}^{K} \sum\limits_{m=1}^{M} \sum\limits_{m'=1, m' \ne m}^{M} \left( {h}^n_{m,k,i} (t) \right)^{*} {h}^n_{m',k, i} (t) {g}^n_{m', i} (\boldsymbol{\theta}_t)}_{\text{\normalfont interference term}} \nonumber\\
& \quad + \underbrace{\sum\limits_{m=1}^{M} \left( \frac{1}{K} \sum\limits_{k=1}^{K} \left( {h}^n_{m,k, i} (t) \right)^{*} \right) z_{k,i}^n (t)}_{\text{\normalfont noise term}}. 
\end{align}
There are three terms with ${y}^n_i (t)$ specified by signal, interference, and noise terms, respectively, in \eqref{ReceivedVectorPSScheCombAntennasReWrith}. With the law of large numbers, as the number of antennas at the PS $K \to \infty$, the signal term approaches 
\begin{align}\label{SigTermApproaches}
y_{i, {\rm{sig}}}^n (t) \triangleq \alpha_t \sigma_h^2 \sum\nolimits_{m=1}^{M} {g}^n_{m,i} (\boldsymbol{\theta}_t), \quad i \in [s], n \in [N],
\end{align}
from which the PS can recover 
\begin{subequations}\label{PSSigTermRecReIm}
\begin{align}\label{PSSigTermRecRe}
\frac{1}{M} \sum\nolimits_{m=1}^{M} g_{m,2(n-1)s+i} \left(\boldsymbol{\theta}_{t}  \right) &= \frac{ {\rm{Re}} \left\{ y_{i, {\rm{sig}}}^n (t) \right\} }{\alpha_t M \sigma_h^2},\\
\frac{1}{M} \sum\nolimits_{m=1}^{M} g_{m,(2n-1)s+i} \left(\boldsymbol{\theta}_{t}  \right) &= \frac{ {\rm{Im}} \left\{ y_{i, {\rm{sig}}}^n (t) \right\} }{\alpha_t M \sigma_h^2}.\label{PSSigTermRecIm}
\end{align}
\end{subequations}
However, the interference term in \eqref{ReceivedVectorPSScheCombAntennasReWrith} does not allow the exact recoveries of $\frac{1}{M} \sum\nolimits_{m=1}^{M} g_{m,2(n-1)s+i} \left(\boldsymbol{\theta}_{t}  \right)$ and $\frac{1}{M} \sum\nolimits_{m=1}^{M} g_{m,(2n-1)s+i} \left(\boldsymbol{\theta}_{t}  \right)$ from ${y}^n_i (t)$, which is observed at the PS. To analyze the interference term, we first define, for $i \in [s]$, $n \in [N]$, 
\begin{align}\label{IntTerAnDef}
\mathfrak{h}_i^n (t) \triangleq \frac{1}{K} \sum\limits_{k=1}^{K} \sum\limits_{m=1}^{M} \sum\limits_{m'=1, m' \ne m}^{M} \left( {h}^n_{m,k,i} (t) \right)^{*} {h}^n_{m',k, i} (t). 
\end{align}
It is easy to verify that the mean and the variance of $\mathfrak{h}_i^n (t)$ are given by
\begin{subequations}\label{MeanVarMathFrakh}
\begin{align}\label{MeanMathFrakh}
\mathbb{E} \left[ \mathfrak{h}_i^n (t) \right] =& 0,\\ 
\mathbb{E} \left[ \left| \mathfrak{h}_i^n (t) \right|^2 \right] =& \frac{M(M-1) \sigma_h^4}{K},\label{VarMathFrakh}  
\end{align}
\end{subequations}
respectively. We note that the gradient values computed at each iteration are independent of the channel realizations experienced during the same iteration. Accordingly, by fixing the gradient values, from the analysis in \eqref{MeanVarMathFrakh}, we conclude that the interference term in \eqref{ReceivedVectorPSScheCombAntennasReWrith} has zero mean and a variance that scales with $M^2 / K$. Thus, for a fixed number of wireless devices $M$, the variance of the interference term in \eqref{ReceivedVectorPSScheCombAntennasReWrith} approaches zero as $K \to \infty$. In practice, it is feasible to employ sufficiently large number of antennas at the PS exploiting massive multiple-input multiple-output (MIMO) systems \cite{RusekScaleUpMassiveMIMO}.

\begin{figure*}[t!]
\centering
\begin{subfigure}{.5\textwidth}
  \centering
  \includegraphics[scale=0.61,trim={20pt 7pt 36pt 40pt},clip]{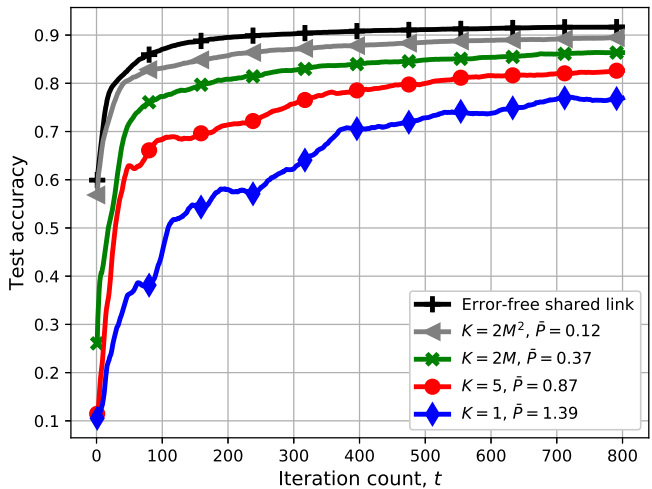}
  \caption{Noise variance, $\sigma_z^2 = 20$}
  \label{FigTestAccNoise10Perfect_2}
\end{subfigure}%
\begin{subfigure}{.5\textwidth}
  \centering
  \includegraphics[scale=0.61,trim={20pt 7pt 36pt 40pt},clip]{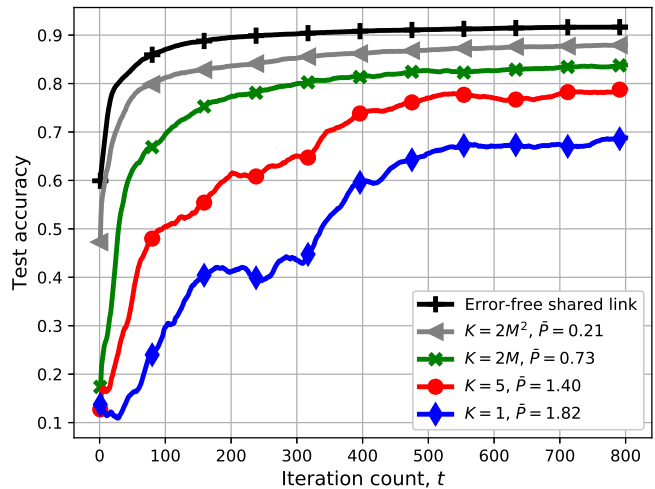}
  \caption{Noise variance, $\sigma_z^2 = 100$}
  \label{FigTestAccNoise50Perfect_2}
\end{subfigure}
\caption{Test accuracy of the proposed multi-antenna analog DSGD algorithm without CSIT for different number of antennas values $\left( K \in \{ 1,5,2M,2M^2 \} \right)$ and noise variances $\sigma_z^2$.}
\label{FigTestAccNoise1050Perfect_2}
\end{figure*}

According to the above analysis, the PS estimates $\frac{1}{M} \sum\nolimits_{m=1}^{M} g_{m,2(n-1)s+i} \left(\boldsymbol{\theta}_{t}  \right)$ and $\frac{1}{M} \sum\nolimits_{m=1}^{M} g_{m,(2n-1)s+i} \left(\boldsymbol{\theta}_{t}  \right)$, for $i \in [s]$, $n \in [N]$, through
\begin{subequations}\label{PSSigTermRecReImEst}
\begin{align}\label{PSSigTermRecReEst}
\hat{g}_{2(n-1)s+i} \left(\boldsymbol{\theta}_{t}  \right) &= \frac{ {\rm{Re}} \left\{ y_{i}^n (t) \right\} }{\alpha_t M \sigma_h^2},\\
\hat{g}_{(2n-1)s+i} \left(\boldsymbol{\theta}_{t}  \right) &= \frac{ {\rm{Im}} \left\{ y_{i}^n (t) \right\} }{\alpha_t M \sigma_h^2},\label{PSSigTermRecImEst}
\end{align}
\end{subequations}
respectively. It then utilizes the estimated vector $\hat{\boldsymbol{g}} (\boldsymbol{\theta}_t) \triangleq \left[ \hat{g}_{1} \left(\boldsymbol{\theta}_{t} \right), \cdots, \hat{g}_{d} \left(\boldsymbol{\theta}_{t} \right) \right]^T$, which can provide a good estimate of the actual average of gradients if a sufficiently large number of PS antennas are employed, to update the model parameters. 

\begin{remark}\label{RemPowerDecay}
We note that with SGD the empirical variances of the gradient estimates decay over time and approach zero asymptotically \cite{BottouLargeScaleSGD,ScalableDNNStorm,DCLimitedPrecisionGupta,MohammadDenizDSGDCS,UseLocalSGDLin}. Thus, for robust communication of the gradient estimates against noise at each iteration of the DSGD algorithm, it is reasonable to increase the power allocation factor $\alpha_t$ over time.        
\end{remark}

\begin{remark}\label{RemCompression}
We remark that the main focus in this paper is to develop techniques to perform a DSGD algorithm at the wireless edge with no CSIT. We propose to employ multiple antennas at the PS, which can help to mitigate the effect of fading, and, in the limit, align the received signals at the PS. We can further employ some of the existing schemes in the literature providing more efficient communication over the limited bandwidth wireless MAC, such as the idea of linear projection proposed in \cite{MohammadDenizDSGDCS}. We leave the analysis of such combined techniques to future work.       
\end{remark}

\section{Numerical Experiments}\label{SecExperiments}
Here we evaluate the performance of the proposed analog DSGD algorithm with no CSI available at the wireless devices. We are particularly interested in investigating the impact of the number of PS antennas on the performance of the proposed scheme. We run experiments on MNIST dataset \cite{LeCunMNIST} with $60000$ training and $10000$ test samples, and train a single layer neural network with $d=7850$ parameters utilizing ADAM optimizer \cite{ADAMDC}. We train the network for $T=800$ iterations. 

We consider $M=20$ wireless devices in the system. To have a realistic model of data distribution across the devices for the wireless edge learning model, we assume that each device has access to $1000$ training data samples selected at random from the training dataset. Thus, some of the training data samples are not assigned to any device, and the data samples across different devices may not be independent. For simplicity, we assume that the $s$ channel gains associated with each OFDM symbol from each device to each PS antenna are i.i.d., and $\sigma_h^2 = 1$. The performance is measured as the accuracy with respect to the test samples based on the updated model parameters at each DSGD iteration.

For numerical comparison, we also consider the benchmark scenario, in which the PS receives the actual average of the gradient estimates $\frac{1}{M} \sum\nolimits_{m=1}^{M} \boldsymbol{g}_m \left(\boldsymbol{\theta}_{t} \right)$, and updates the parameter vector according to this noiseless observation at each DSGD iteration. We refer to this as the error-free shared link scenario, and its accuracy can serve as an upper bound on the performance of the proposed analog DSGD scheme.  

In Fig. \ref{FigTestAccNoise1050Perfect_2} we illustrate the performance of the proposed analog DSGD scheme with no CSIT for different $K$ values and different noise levels. We consider $K \in \{ 1, 5, 2M, 2M^2 \}$, and investigate the performance of the proposed scheme for $\sigma_z^2 = 20$ and $\sigma_z^2 = 100$ in Figures \ref{FigTestAccNoise10Perfect_2} and \ref{FigTestAccNoise50Perfect_2}, respectively. We also include the performance of the error-free shared link scenario. We set the power allocation factor $\alpha_t = 1 + t/1000$, $t \in [T]$, and for simplicity, we assume that $s = d/2$ resulting in $N=1$. We note that, for a fixed power allocation $\alpha_t$, $\forall t$, the value of $s$ does not have any impact on the accuracy of the considered schemes; instead, any change in $s$ scales the average transmit power, whose value is proportional to $N$. As it can be seen, employing more antennas at the PS results in a higher accuracy with the improvement more highlighted when the noise level is higher. This is due to the fact that increasing $K$ mitigates the effects of both the interference and noise terms, inferred from \eqref{ReceivedVectorPSScheCombAntennasReWrith}. Thus, the advantage of having more PS antennas is more pronounced when the channel is noisier. For example, even when $\sigma_z^2 = 100$, the proposed scheme with $K = 2M^2$ PS antennas and average power $\bar{P}=0.21$ provides a slightly smaller accuracy than that of the error-free shared link scenario; this result indicates the success of the proposed scheme in mitigating the noise term even when the ratio $\bar{P}/{\sigma_{z}^2}$ is relatively small. 
We further observe that, compared to having a single-antenna PS, the accuracy improves by exploiting even a few antennas at the PS, e.g., $K=5$, where the improvement is much higher when the channel is noisier, i.e., $\sigma_z^2 = 100$ case. We note that, with all the other parameters fixed, the required average transmit power reduces with $K$, which verifies a faster convergence rate with higher $K$ resulting in a faster reduction in the empirical gradients' variances over time. The same observation is made by reducing $\sigma_z^2$ from $100$ to $20$ while all the other parameters are fixed.

\section{Conclusions}\label{SecConc}
We have studied DSGD at the wireless edge, where wireless devices compute the gradient estimates based on their available limited datasets, and transmit their estimates to the PS over a wireless fading MAC. To make the model more realistic, we have assumed that the devices do not have CSI for the underlying fast fading channel. With the goal of recovering the average gradient estimates at the PS, we have developed an analog DSGD technique, where the effect of fading, which cannot be cancelled at the transmitters due to the lack of CSIT, is alleviated by employing multiple antennas at the PS. Theoretical analysis, corroborated with numerical results, indicates that, with the proposed approach, increasing the number of PS antennas provides a better estimate of the average gradients through a better alignment of the desired signals, as well as elimination of the interference and noise terms. Asymptotically, the proposed DSGD scheme guarantees, despite the lack of CSIT, that the wireless MAC becomes deterministic, and both the fading and noise effects disappear.   

\bibliographystyle{IEEEtran}
\bibliography{Report}

\end{document}